\documentclass[times,12pt]{article}
\pdfoutput=1
\usepackage{amsmath,amssymb,amsfonts,latexsym,amsthm,enumerate,url}
\usepackage{graphicx}
\usepackage{hyperref}

\newcommand{\beq}[1]{\begin{equation}\label{#1}}
\newcommand{\enq}[0]{\end{equation}}

\newcommand{\remove}[1]{}

\newcommand{\comment}[1]{}

\begin{document}

\title {Google's Quantum Supremacy Claim: Data, Documentation, and Discussion}
\author {Gil Kalai, Yosef Rinott, and Tomer Shoham}
\maketitle

\begin{abstract}

In October 2019, {\it Nature} published a paper \cite {Aru+19} describing 
an experiment that took place at Google.
The paper claims to demonstrate quantum (computational) supremacy on a 53-qubit quantum computer. Since September 2019 we 
have been involved in a long-term project to study various statistical
aspects of the Google experiment. We have been trying to gather the
relevant data and information in order to reconstruct and verify those parts of
the Google experiment that are based on classical computations (except when the required computation is too heavy), and to perform a statistical analysis on the data.  
This document describes the available data and information for the Google experiment, some main questions in the evaluation of the experiment, and some of our results and plans. 

\end{abstract}




\section {Introduction}

The 2019 paper ``Quantum supremacy using a programmable
superconducting processor'' \cite {Aru+19} claimed that
Google's  Sycamore quantum computer of 53 qubits and depth
20, performed a certain computation in about 200 seconds, while
a state-of-the-art classical supercomputer
would take, according to the Google team's estimate,
approximately 10,000 years to perform the same computation. 
Google's Sycamore quantum computer performed a \textit{sampling task}; that is,
it generated random bitstrings of length 53,
with considerable noise,
from a certain discrete probability distribution supported on all such $2^{53}$ bitstrings. 
(Here, a {\it bitstring} refers to a vector with 0,1 entries.)
The specific sampling task performed by Google is referred to as random circuit sampling (RCS, for short).
Google's announcement of  quantum supremacy was compared by
various writers (see, e.g., \cite {Aar19}) to landmark technological
achievements such as the Wright brothers'
invention of a motor-operated airplane, the launching of Sputnik, and the
landing of humans on the moon, 
as well as to landmark scientific achievements
such as Fermi's demonstration of a nuclear chain reaction,
the discovery of the Higgs boson, and the LIGO detection of gravitational waves.


In 2020  
a team
from the University of Science and Technology of China (USTC) claimed \cite{Zho+20} that the sampling task
computed by their photonic {\it Jiuzhang} quantum computer  
would take 2.5 billion years
to perform on a classical supercomputer. USTC's quantum
computer took about 200 seconds to complete the task. This task is referred to as
Gaussian boson sampling (GBS, for short).
In 2021, another team from USTC repeated the
Google RCS experiment with their 
{\it Zuchongzhi} quantum computer of
60 qubits and depth 24  \cite {Wu+21,Zhu+22}, and claimed to achieve
an even stronger form of quantum advantage compared to the Google experiment. 

The Google experiment represented a very large leap 
in various aspects of the human ability to control
noisy quantum systems. For example, the previous experiment reported by the Google AI team used only nine qubits \cite {Nei+18}. 
This leap is especially impressive in terms of the
dimensions of the Hilbert space representing a state of the computer 
(from 100--500 dimensions in \cite{Nei+18} to $10^{16}$ dimensions in \cite {Aru+19}.)

Google's quantum supremacy claim is based on two ingredients. 
The first ingredient is an assertion about the quality of the samples produced by the quantum computer. This quality is described in terms of an important parameter called the fidelity. The second ingredient is an assertion about the difficulty of achieving samples of the same quality by a classical supercomputer. The second assertion, and with it the entire quantum supremacy claim, was largely refuted by research of several groups \cite{PanZha21,PCZ22,ZSW20,KPY21,KPZY21,PGN+19,Gao+21} (among others) 
that exhibited classical algorithms that are ten 
orders of magnitude faster than those used in the Google paper. 
This was achieved, for example, by Pan, Chen, and Zhang \cite {PCZ22} in 2021. 
Our study 
concentrates on the first claim dealing with the estimated fidelity of Google's samples.

As we have already mentioned, the Google announcement was regarded as a major scientific and technological breakthrough.
On its own it gave some evidence that the ``strong Church--Turing thesis'' had been violated and it was described
as an ironclad refutation of claims by some scientists (including Kalai)
that quantum computation is not possible.  
Of no less importance is the fact that quantum supremacy was considered as a major intermediate step toward
exhibiting experimentally quantum error-correction codes needed for building larger quantum computers.
The announcement of quantum supremacy stirred a great deal of enthusiasm among scientists and in the general public, and
garnered significant media attention. It had a substantial impact;  
for example, following the media attention surrounding the leaking of the quantum supremacy claims 
around September 22, 2019, the value
of bitcoin (and other digital currencies) sharply dropped by more than 10\% and it is a reasonable
possibility 
that given the
potential threat of quantum computers
to the safety of digital currencies, the quantum supremacy claims caused 
this drop.

\begin{figure}
\centering
\includegraphics[scale=0.6]{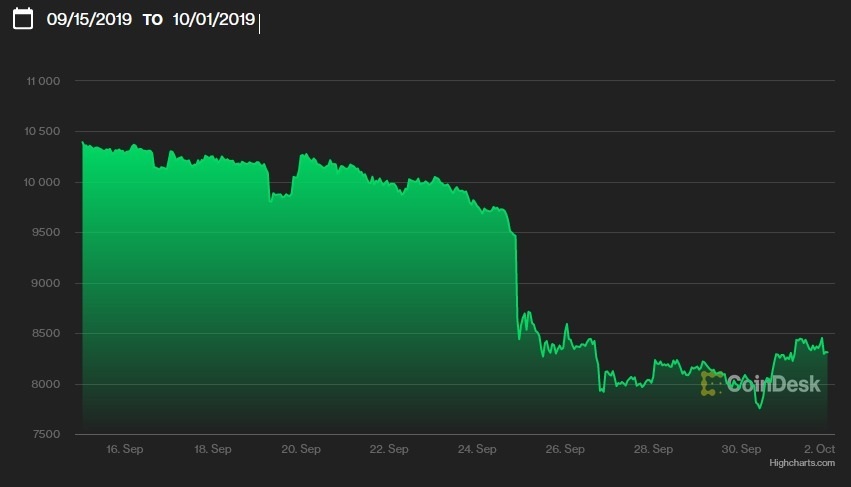}
\caption{{\it The price of bitcoin in USD in a period of 4 weeks around 9/23/2019. Source: CoinDesk}}
\label{fig:1}
\end{figure}

\subsection* {Scrutinizing the Google supremacy claim}

The Google paper was briefly posted on a NASA server and
became publicly available a month before it was published in October 2019.
Following the announcement of the quantum supremacy claim, the first-named author (and other researchers) raised various concerns about some aspects of the claims. 
A few months later the authors initiated what has become a   
long-term project to study various statistical aspects of the Google experiment.  
In particular, we have been trying to gather the relevant data and information and
to reconstruct and verify those parts of the
Google experiment
that are based on classical computations, except when the required computation is too heavy.
(We carried out some heavy computation on the cloud for which we put a cap of 2000 dollars on our spending.) We also performed several ``sanity tests'' of the experiment. 
\subsection *{The structure of this paper}
In Section 2 we provide a brief background on the Google experiment and describe the various types of circuits used in the experiment. We also present the chronology of the various Google experiments performed on the Sycamore quantum computer leading to the ultimate quantum supremacy experiment,  
as reported to us by the Google team. In Section 3 we describe the nature of the calibration process. In Section 4 we describe the data requested from the Google team, the data that was provided between October 2019 and June 2022, and some other details related to the Google experiment and other NISQ quantum supremacy experiments. In Section 5 we present two proposals for future experiments. In Section 6 we discuss what we regard as the main questions in the evaluation of the Google experiment and list some confirmations, refutations, concerns, and weaknesses, and in Section 7 we briefly discuss where we are now in our study.   
\comment{
\subsection* {Two concrete questions}

Let us raise two motivating questions about
(verifiable full) random circuit sampling of a circuit $C$ of the kind discussed in the
Google paper with $n=22$ qubits and depth $m=14$. (We will come back to these questions
and add a third one in Section \ref{s:3que}.) 

\begin {enumerate}

\item
  Can humanity produce at present samples from a quantum circuit $C$
  which are good approximation of the Google noise model
  or any other specific noise model?

\item
Did humanity reach the ability to produce samples
for quantum circuit $C$  with linear cross entropy fidelity estimated above 0.15?

\end {enumerate}

In both questions we refer, of course, to genuine quantum device and not to classical simulations. 
}

\section {Google's quantum supremacy claim}
\subsection {A brief background}
\label {s:g1}
In this paper we will assume knowledge of the Google experiment,
Google's noise model,
Google's ${\mathcal F}_{XEB}$ linear cross-entropy fidelity estimator,
and Google's Formula (77) in \cite {Aru+19S} for predicting the fidelity
of a circuit from the fidelity
of its components. We will give here a brief summary of these topics.

The Google experiment is based on the building of a quantum computer (circuit), 
with $n$ superconducting qubits,
that performs $m$ rounds of computation.
The computation is carried out by 1-qubit and 2-qubit gates. 
At the end of the computation the qubits are measured, leading 
to a string of zeroes and ones of length $n$. 
The ultimate experiment (on which Google's central claim was based) was for $n=53$ and $m=20$. It involved
1113 1-qubit gates and 430 2-qubit gates. For that experiment  
the Google team produced a sample of three million 0-1 vectors of length 53. 

Every circuit $C$ with $n$ qubits 
describes a probability distribution ${\bf P}_C(x)$ for 0-1 vectors of length $n$. 
(In fact, it describes a $2^n$-dimensional vector of complex amplitudes; for every 0-1 vector
$x$, there is an associated amplitude $z(x)$ and ${\bf P}_C(x)=|z(x)|^2.$)
The quantum computer enables sampling according to the probability distribution ${\bf P}_C(x)$
with a considerable amount of noise. When $n$ and $m$ are not too large, classical simulations enable computation of the amplitudes themselves (and hence the probabilities ${\bf P}_C(x)$). Google's quantum supremacy claim is based on the fact that these classical simulations quickly become infeasible as $n$ and $m$ grow. 

Google's basic noise model for the noisy samples produced by their quantum computer is 
\begin {equation}
\label{e:gnm}
{\bf N}_C(x) =  \phi {\bf P}_C + (1-\phi) 2^{-n},
\end {equation}
where $\phi$ is the {\it fidelity}, a parameter that roughly describes the quality of the sample. (The fidelity has a precise meaning in terms of the actual noisy quantum process carried out by the quantum computer.)

Based on their noise model (and the fact that the distribution ${\bf P}_C$ is an instance of a Porter--Thomas distribution), the Google paper describes a statistic called the linear cross-entropy estimator (denoted by ${\cal F}_{XEB}$.)
Once the quantum computer produces a sequence ${\bf \tilde x}$ of $N$ 
samples ${\bf \tilde x}= ({\bf \tilde x}^{(1)},{\bf \tilde x}^{(2)},\dots, {\bf \tilde x}^{(N)})$, the following ``linear cross-entropy" ${\cal F}_{XEB}$ estimator of the fidelity is computed:

\begin {equation}
\label {e:fxeb}
{\cal F}_{XEB}({\bf \tilde x})=\frac{1}{N}\sum _{i=1}^N 2^n{\bf P}_C({\bf \tilde x}^{(i)}) -1.
\end {equation}
Computing ${\cal F}_{XEB}$ requires knowledge of ${\bf P}_C(x)$ for sampled bitstrings. 

The Google quantum supremacy claim is based also on the following a priori prediction of the fidelity of a circuit based on the probabilities of error for the individual components:

\begin {equation}
\label {e:77}
~\hat  \phi~=~ \prod_{g \in {\cal G}_1} (1-e_g) \prod_{g \in {\cal G}_2} (1-e_g) \prod_{q \in {\cal Q}} (1-e_q).
\end {equation}

Here ${\cal G}_1$ is the set of 1-gates, 
${\cal G}_2$ is 
the set of 2-gates, 
and
${\cal Q}$ is the set of qubits. For a gate $g$, the term $e_g$ in the 
formula refers to the probability of an error (1 minus the fidelity) of the individual gate $g$. 
For a qubit $q$, $e_q$ is the probability of a readout error when we measure the qubit $q$.

The Google supremacy paper \cite {Aru+19} made two crucial claims regarding the ultimate 53-qubit sample experiment.

\begin {itemize}
\item[A)] The fidelity $\phi$ of their sample is above $1/1000$. 
 

\item [B)] Producing a sample with similar fidelity 
would take 10,000 years on a classical supercomputer.
\end {itemize}
As for claim A) regarding the value of $\phi$, 
the argument  relies on an 
extrapolation argument that has two ingredients. 
One ingredient is a few hundred experiments in the classically tractable regime: the 
regime where the  probability distribution ${\bf P}_C$ can be
computed by a classical supercomputer and the performance of the 
quantum computer can be tested directly. The other
ingredient is the theoretical Formula (\ref {e:77}) for predicting the fidelity.
According to the paper, the fidelity of 
entire circuits closely agrees with the prediction of Formula (\ref {e:77}) 
(Formula (77) in \cite {Aru+19S}) 
with a deviation below 10--20 percent. 
There are around 200 reported experiments in the
classically tractable regime including ones carried out on simplified
circuits (which are easier to simulate on classical computers). These experiments
support the claim that the
prediction of Formula (77) for the fidelity is indeed very robust and applies
to the 53-qubit circuit in the supremacy regime.

As for claim B) regarding the classical difficulty, the Google team
mainly relies on extrapolation from the running time of a
specific algorithm they used. They also rely on the computational complexity 
support for the assertion
that the task at hand is asymptotically difficult.

{\bf Remark:} The Google team proposed a pretty good approximation of Formula (\ref {e:77}) based on averaged fidelities: 

\begin {equation}
\label {e:77s}
{\hat \phi}^*= (1-0.0016)^{|{\cal G}_1|} (1-0.0062)^{|{\cal G}_2|} (1-0.038)^n .
\end {equation}


\subsection {The types of circuits of the Google experiment} 

The circuits used in the Google experiment had the following structure. 
The qubits were arranged on a planar grid, and so a single qubit was identified via two 
coordinates, like qubit $(3,3)$. The circuits had two types of layers:
one type of layer consists of 2-gates acting on pairs of (neighboring) qubits. After each such layer of 2-gates there was another layer of randomly chosen 1-gates acting on every qubit. The layers of 1-gates consist of the ``programmable" ingredient in the experiment and they change from circuit to circuit. The layers of 2-gates are fixed throughout the experiment according to a certain pattern. The pattern {\bf EFGH} was used in all experiments conducted between February and May 2019    
and a new pattern {\bf ABCDCDAB} was used in June 2019 to produce the samples of the ``supremacy" circuit, namely, circuits that would require heavy classical computation. Each letter like {\bf E} corresponds to a fixed set of 2-gates acting in parallel on the qubits 
(For circuits with a smaller number of qubits we regard only 2-gates that involve the qubits in the circuit). By pattern {\bf EFGH} we mean that we first apply a layer of random 1-gates on all qubits, then apply 2-gates according to {\bf E}, then apply another layer of random 1-gates on all qubits, then apply 2-gates according to {\bf F}, and so on  (in a periodic manner). The new pattern {\bf ABCDCDAB} is based on new types of layers for the 2-gates and on a period of length eight. 

The depth $m$ of a circuit refers to the number of layers of 2-gates. For example, the layers of 2-gates of a circuit with pattern {\bf EFGH} and depth $m=14$ are {\bf E, F, G, H, E, F, G, H, E, F, G, H, E, F}.

\subsection* {The six types of circuits} 

The Google experiment relies on six types of circuits. The first three types are: 
\begin {itemize}
  \item [a)]
    The full circuits with pattern {\bf EFGH}. 

  \item [b)]
    The elided circuits with pattern {\bf EFGH}.
  \item[c)]
    The patch circuits with pattern {\bf EFGH}.

\end {itemize}    
Types b) and c) are simplified forms of type a) for which there are quick algorithms to compute the 
amplitudes for all values of $n$ up to 53. For the elided and patch circuits certain 2-gates of the full circuits are removed.  
For patch circuits, all 2-gates between two disjoint sets of qubits are removed and therefore a patch circuit consists of two separate
circuits on these two disjoint sets of qubits. For elided circuits only some of these 2-gates 
are removed. The Google paper \cite {Aru+19} is based on running the quantum computer on a
few hundred circuits of types a), b), and c) and computing the 
fidelity estimator based on the amplitudes.
We note that in the Google experiment all of these different types of circuits are based on the same random 1-gates. Namely, the $i$th files of the full, elided, and patch circuits all have the same random 1-gates. (This remark applies also to the circuits with pattern {\bf ABCDCDAB} considered below.) We also note that the random 1-gates are the same for different values of $n$; namely, the random choices for $n=53$ are applied to every $n<53$.  (The same random 1-gates were used for the $n=53$, $m=14$ experiments with patterns {\bf EFGH} and {\bf ABCDCDAB}; for the later pattern, random gates for the $i$th experiment for 1-gates in level $k$ are the same for all depths $m=12,14,\dots,20$.)   

There was also a preliminary stage of calibration that, based on experiments 
on 1-qubit and 2-qubit circuits, determined the precise
adjustments for the experimental 2-gates.
Those adjustments are the same for all circuits of types a), b), and c). See Section \ref {s:cal}.

As far as we understand, the initial plan of the Google team to demonstrate 
quantum supremacy   
was to compute the empirical fidelities for circuits of type b) and c)
for a number of qubits between 12 and 53 and for circuits of
types a) for a number of qubits up to $n=43$ and to use this
information to estimate the fidelity for circuits of type
a) with 53 qubits and depths ($m$) between 12 and 20.  
However, in May 2019, the Google team discovered 
an efficient
classical algorithm for the circuits of type a). 
(Because of this discovery, the full circuits with pattern {\bf EFGH} are also referred to as ``verifiable full circuits.'')
The new algorithm discovered by the Google team is related to tensor networks methods, which later
led to important discoveries in this direction.\footnote {The Google team pointed out that they also implemented a related tensor network method in 2017 \cite{BISN17}.}  
Using this algorithm, the Google team computed the amplitudes for the
experimental bitstrings for type a) circuits with $n=53$ and $m=14$.

Subsequently, for the purpose of demonstrating quantum supremacy, 
the Google team moved in May 2019 to a different
architecture based on a new pattern {\bf ABCDCDAB} 
that is harder to simulate with classical supercomputers. This led to the following three new types of circuits:

\begin {itemize}
  
\item [d)] The full circuits with pattern {\bf ABCDCDAB}. (These circuits are also referred to as ``supremacy full circuits.'') 
\item [e)]
    The elided circuits with pattern {\bf ABCDCDAB}.
\item[f)]
    The patch circuits with pattern {\bf ABCDCDAB}.
\end {itemize}

For this architecture the Google team produced samples only with number of qubits $n=53$ and depth $m=12,14,16, 18$, and $20$.
The simulators available to the Google
team were not powerful enough to compute the amplitudes and to check the fidelity of 
circuits of type d).

The circuits with pattern {\bf ABCDCDAB} required the calibration
process to be run again on 1-qubit and 2-qubit circuits based on the new architecture and this 
led to new adjustments for the 2-gates.
Circuits with the new pattern {\bf ABCDCDAB}, as well as the calibration process applied to them, were not tested
for less than 53 qubits.

\subsection {A brief chronology of the Google quantum supremacy experiments}
\label {s:chr}

Prior to the development of the Sycamore quantum computer,
in 2018 the Google team had attempted to develop a 72-qubit chip
called
``Bristlecone,'' but due to difficulties the team later proceeded to the development of the 54-qubit Sycamore quantum 
computer with 53 effective qubits (due to the malfunctioning of one of the qubits).  
Prior to the experiments reported in the Google paper there were five earlier experiments that took place in February and March 2019. All these experiments are based on full circuits of pattern {\bf EFGH}.\footnote {Some more details can be found in versions 1 and 2 of this paper on the arXiv.} 

The Google paper is based on two experiments. The first experiment with  circuits of pattern {\bf EFGH} started April 22, 2019, and is based on patch, elided, and full circuits with depth $m=14$ and number of qubits $n=12,14, \dots 38$ and all integers between $39$ and $53$.
The second experiment with circuits of pattern {\bf ABCDCDAB} started June 13, 2019, and is based on patch, elided, and full circuits  with depths 
$m=12,14,16,18,20$, and number of qubits $n=53$. 
Each experiment took 1--2 days to perform.

\section {The nature of the calibration process}
\label {s:cal}
Google's Sycamore quantum computer has systematic deviations from the ideal circuit it describes, and 
the calibration process accounts for small systematic errors in the experimental circuits
compared to the random circuits they represent. The calibration process is crucial 
since the combined effect of such systematic errors can slash the fidelity to zero and it is important to account for them.

By calibration process we refer in this paper to a method that,
based on multiple runs of
1-qubit and 2-qubit quantum circuits,
adjusted the definition of
the 2-gates of the experimental circuits so that certain systematic
forms of noise would be greatly reduced. This calibration adjustment was carried out simultaneously in
all the experiments for patch, elided, and verifiable full circuits.

It is easy to identify from the Python program how the random circuits were modified by the calibration.
(We could also use the QSIM files for that purpose.)
For example, consider a single  2-gate acting on two qubits $A=(3,3)$ and $B=(3,4)$.

\subsubsection *{The ideal description of the circuit}
In the ideal description of the circuit, this 2-gate is described in the Python program for the circuit as follows:

\medskip

{\tt
\noindent
{\tt
\noindent
cirq.Moment(

            
                cirq.FSimGate(theta=1.57079632679, phi=0.52359877559).
                
                on(
                    cirq.GridQubit(3, 3), cirq.GridQubit(3, 4)),
}
}
\medskip

\noindent
In this 2-gate (which is sometimes referred to as the ``standard 2-gate") 
$\theta = \pi/2$, and $\phi = \pi/6$.
We will now show the part of the program that describes 2-gate adjustments to this specific gate. 
The calibration consists of adding to the description of the circuits  
two pairs of fixed rotations on each of the 2 qubits
involved in this particular 2-gate. (One pair of rotations acts on the two qubits before executing the 2-gate itself and another pair of rotations acts after executing the 2-gate.)

\subsubsection *{The added single-qubit rotations}

The two added single-qubit rotations acting before the 2-gate is executed are

\medskip

{\tt
\noindent
cirq.Rz (np.pi * 2.5333591271878086). on (cirq.GridQubit (3, 3)), 
\noindent
cirq.Rz (np.pi * -2.4748096263683066). on (cirq.GridQubit (3, 4)).
}

\medskip

\noindent
The same 2-gate involving qubits (3,3) and (3,4) occurs several times in a circuit and the single-qubit rotations are different in each of these occurrences. These two pairs of added single-qubit rotations are the same for the $k$-th appearance of this gate for all the 
circuits.
The values that appear here in the single-qubit rotations, as well as the adjustments for the definition of 2-gates, were computed based on many experiments for 1-qubit and 2-qubit circuits for these two qubits (3,3) and (3,4). We have neither the precise algorithm for this computation nor the data with the outcomes of the 1-qubit and 2-qubit experiments. See Remark 1 for further details.  


\subsubsection *{The 2-gate adjustment}
In addition to two pairs of single-qubit rotations,  the 2-gate acting on qubits $A$ and $B$ is modified to
\medskip

{\tt
\noindent
cirq.Moment(

            
                cirq.FSimGate(theta=1.2947043217999283, phi=0.4859467238431821).
                
                on(
                    cirq.GridQubit(3, 3), cirq.GridQubit(3, 4)),
}

\bigskip
            
            The calibrated values $\theta=1.2947043217999283$ and $\phi=0.4859467238431821$ are the
same for all 2-gates acting on these two qubits in all circuits. (Sometimes the calibrated 2-gate is referred to as the ``native" 2-gate.) 
These values for $\theta$ and $\phi$ replace the values for the ``standard" gate in the original random circuit $\theta = \pi/2$ and $\phi = \pi/6$, respectively.



\subsubsection *{Further details and remarks} 

1. We  note that the adjustments for the 2-gate involving qubits $A$ and $B$ depend
(only) on multiple runs of 1-qubit and 2-qubit quantum circuits involving these two qubits. 
These multiple runs were not conducted in isolation but in parallel, and each run represented  layers of 2-gates of the same type ({\bf E},{\bf F},{\bf G}, or {\bf H} for the full verifiable circuits or {\bf A}, {\bf B}, {\bf C}, or {\bf D} for the full supremacy circuits), where, as in the experimental circuits, between the layers there were random 1-gates. Note that in applying layers of 2-gates of the same type along with random 1-gates between the layers, we interact every qubit with (at most) a single other qubit, to obtain many 2-qubit circuits running in parallel.  

2. We also note that there might have been additional actions to account for
the fact that certain components of the Sycamore chip degrade over time.

3. In an October 2019
Internet discussion \cite {SO:Gidney2}, 
Craig Gidney from the Google team asserted  that using Sycamore's standard gates instead of native gates halves the fidelity. 
(A similar assertion, with reference to Figure S30 in \cite {Aru+19S}, was made a year later by Sergio Boixo \cite {SO:Boixo}.)
However, this assertion referred only to the two-gate adjustments and
not to the one-gate adjustments, and moving to the original circuit with standard gates slashes the fidelity to zero. 
(The effect of the 2-gate adjustments represents the worst-case situation for $n=53, m=20$, 
and it is smaller for smaller circuits.)
Boixo later clarified 
that for later experiments conducted by Google in 2020, the Google team developed a
method for carrying out the 1-qubit calibrations using physical control of the device rather
than using a modification of the definition of the circuit.
(This later development is not relevant to Google's quantum supremacy 
experiment and we did not study it.)

4. The calibration method was based on running full circuits having only one type of layer of 2-gates, with layers of random 1-gates between them, and making the adjustments for 2-gates so as to maximize the ${\cal F}_{XEB}$ value. Note that circuits that have only one type of layer of 2-gates can be divided into noninteracting 2-qubit circuits and 1-qubit circuits.
It seems that the same algorithm 
can be applied also when the layers of 2-gates are not the same and, in particular, the algorithm can be applied to the experimental circuits themselves.

5. The same 2-gate occurs several times in a circuit. We note that the 2-gate adjustments are the same for all these occurrences, but the added single-qubit rotations are different for different occurrences of the same 2-gate in a circuit.

\section {Data, documentations, and discussions}
\label {s:ddd}

\subsection {The supplementary data}

The Google paper was leaked around September 23, 2019, and published by {\it Nature} a month later on October 23, 2019. The data for the Google experiment can be found in \cite{Aru+19D}. 
The Google team uploaded the data to the server on the following five dates:

\subsubsection * {\bf October 22/23, 2019, publication date}

1. Files with 500,000 bitstrings for many of the experimental circuits with $n<53$ were uploaded and for circuits with $n=53$ files with a few million bitstrings were uploaded.
The bitstrings for several numbers of qubits $n$ were missing on this date (e.g., missing for $n$ = 16, 32, 34, 38, 39, and all integers between $42$ and $51$). In addition, the bitstrings for the patch circuits were missing. 

\noindent
2. QSIM files with descriptions for the corresponding experimental circuits, 
and Python files for computing their amplitudes, were uploaded.


\subsubsection *{\bf January 23, 2020}
Missing bitstrings for full and elided circuits were uploaded. 
Amplitude files for all but roughly 200 circuits (not including patch circuits) were uploaded.  
Some additional data including information on noise for individual components of Google's  Sycamore quantum computer was uploaded.

\subsubsection * {\bf January 22, 2021 \& May 18, 2021}
Amplitude files for all but eight circuits (not including patch circuits) for which the amplitudes were computed in J\"ulich Research Center, Germany, were uploaded. Readout error data was uploaded.

\subsubsection * {\bf June 13, 2022}
The J\"ulich amplitudes were uploaded. Bitstring files (100,000 bitstrings per circuit) and QSIM and Python files for the patch circuits with $n<53$ were uploaded. (These files are still missing for $n=53$, pattern {\bf EFGH}.)

\subsection {Requests for data}
\label {s:req}

\subsubsection {Early requests}

Here are Kalai's
requests for data from the Google team from early October 2019 (items (a)--(e)) and November 2019 (item (f)). 
\begin {enumerate}

\item [(a)]  The bitstrings produced by the quantum computer and a description of the circuits.
\item [(b)]  The amplitudes computed for the samples for each circuit.
\item [(c)]  The full list of amplitudes. (Here we refer to the $2^n$ amplitudes for all bitstrings and not only for those in the sample.) 
\item [(d)]  Larger samples of the quantum circuits and a comparison of the empirical distribution with the model.
\item [(e)]  Timetable of the experiments and calibrations.
\item [(f)]  The values of the individual fidelities $e_q$ and $e_g$ for every qubit $q$ and gate $g$ used in Formula (\ref{e:77}). 
\end {enumerate}

\subsubsection {Subsequent requests}
\begin {enumerate}
\item [(g)] Readout error information. We discussed this matter with the Google team in January 2021 and we
  received useful data on the readout errors shortly afterwards.
  In this case, we received the requested data quickly, in a very satisfactory way.

\item [(h)]
  Amplitudes for the verifiable experiments.  As we have already mentioned, the Google team developed  useful algorithms and programs to compute the 
  amplitudes for the experiments of full circuits with pattern {\bf EFGH}.
  (This was the reason for changing to a different pattern for the ultimate experiment.)
  They used these programs for 10 53-qubit depth-14 circuits and the computation 
  required several hours for each circuit
  on the Google (classical) supercomputer. For a useful reliability control
  of the experiment we proposed (in September 2021) that the amplitudes be computed for the remaining (over 100) full circuits, and for at least 2 specific ones.  

\item [(i)]
  The calibration programs for moving from data on 1-qubit and 2-qubit
  circuits to 1-gate and 2-gate adjustments in the definition of the circuits,
  and the raw data gathered from 1-qubit and 2-qubit circuits. 
  We requested these in January 2022.

\end {enumerate}

\subsection {Data provided by the Google team}
\label {s:supply}

\begin {enumerate} 

\item
Google’s researchers were under a press embargo and did not respond to most issues before
the {\it Nature} paper appeared. 

\item
As we mentioned, on October 23, 2019, the paper was published in {\it Nature}, and the supplementary raw data contains request (a)
(bitstrings and a description of the circuits in the form of a Python program for computing the amplitudes)
for many of the circuits used in the experiments. In January 2020 bitstrings for the remaining full and elided circuits were added. 
The data for the patch circuits was uploaded to the server in June 2022 
except for the 53-qubit depth-14 {\bf EFGH} patch circuits.
As we mentioned, the data has only 100K samples per patch circuit. (The Google team also provided a useful 
Python program for splitting each of the patch circuits into two patches.)  

\item
The Google team promised to supply the amplitudes (that they computed as part of the experiment) 
for their samples (request (b)). They 
uploaded the amplitude files for many circuits in
January 2020, and for two-hundred additional circuits in May 2021.
Uploading the amplitudes that were computed for 8 circuits in an external facility (J\"ulich) 
required the approval of the J\"ulich team and they were uploaded in June 2022.  

\item	 Regarding request (c), 
the Google team informed us that the full lists of amplitudes were discarded. (Indeed,
for large values of $n$ this is a huge amount of data.) Moreover, for circuits with pattern {\bf EFGH} the Google team
developed (around May 2019) algorithms for computing the amplitudes for the 
sample that did not
require computing all the $2^n$ amplitudes and these algorithms were used in some of these circuits. 

\item	 Regarding our request (d),  
the Google team did not supply any additional samples produced by the quantum computer, and referred 
to a study of the empirical distribution in an earlier experiment (on 9 qubits) \cite {Nei+18}. 

\item   Regarding the timetable and some details for the experiments and 
calibrations (request (e)),
  in May 2022 we
  received a brief timetable and some useful details on the final experiments and five earlier ones.
  See Section \ref {s:chr}. 
\item
So far we have not received the details regarding individual gate- and qubit- fidelities (request (f)) and the (related) data represented in Figure 2 of Google's paper \cite {Aru+19}. We were also not informed by the Google team whether this data can be shared or not, and we hope that the Google team will share it in the future. (The Google team provided useful approximate versions for Formula (77), such as Formula \eqref {e:77s}, but  recently it turned out that this data is crucial for the study of patch circuits and therefore we have raised this request again.) 

\item 
 Regarding request (h), the Google team told us that they would 
  not provide amplitudes for the verifiable experiments 
  as these computations would require a lot of human and computational effort.

  However, using new algorithms, amplitudes for all the verifiable experiments were computed by
  Kalachev, 
  Panteleev, and 
  Yung \cite {KPY21}
  and the results lend strong support to Google's predictions 
 on the linear cross-entropy fidelity estimates of their samples. 

\item
  Regarding request (i), the Google team informed us that they could not share
  the full proprietary calibration system (which took many years to develop)
  needed to move from data on 1-qubit and 2-qubit to 1-gate and 2-gate adjustments in the definition of the circuits.
  They pointed out that the main innovation of the calibration program had been made public. 


\end {enumerate}


\subsection {The discussion with the Google team}

From October 2019 to December 2022, we had good discussions (initiated by Scott Aaronson), mainly by
email, with the Google team and especially with John Martinis and Sergio Boixo,
on various aspects of their
experiment and on some of our findings. Overall, the Google team welcomed
us (and others) subjecting their experiment to
careful scrutiny, even though they had been aware since 2019 of the first author's concerns 
\cite {Ira19} about the reliability
of the Google quantum supremacy experiment. The discussions were easygoing and congenial.
A video discussion between Boixo, Kalai, Rinott, and Shoham in October 2021 was especially fruitful.
The basic methodology of trying to subject the data from the Google experiment to various ``sanity tests''
was overall agreed upon by the Google team although at times
we had different interpretations of specific findings. 
There was a single issue regarding our concerns about the calibration process
in connection with the 2019 Google video  \cite {Google:video19},
discussed in late 2021, that led to a somewhat more tense exchange, 
where John Martinis criticized attempts to pass judgment on the
Google experiment based on a short video meant for a general audience,
rather than on the paper itself. 
Still, even in this case, we had a useful discussion that ultimately
shed some light on aspects of the Google experiment.

Overall, we were not shy to ask for data and information and, in a few cases,
the Google team was not shy to decline
our requests. Most of our requests for data and
information were met, although, arguably, most of our requests should have been part of
the supplementary material of the Google paper to start with. 

From December 2020 to November 2021, we conducted in parallel a useful 
discussion with Chao-Yang Lu and members of the 
Gaussian boson sampling team from USTC along with members of 
the Google team and other researchers 
regarding the issue of spoofing and $k$-point correlations of the Gaussian boson sampling experiment. (This discussion was also initiated by Scott Aaronson.)    
We also had a brief email correspondence with the USTC team that replicated the Google experiment.


\subsection {The discrepancy with the outcomes of the J\"ulich Research Center team}

One little mystery that was settled following our discussions with the Google team
involved the data from the J\"ulich team. The amplitudes for 8 large circuits with $39, 42$, and $43$ qubits were computed by researchers from the J\"ulich Research Center
using their own powerful simulators 
and (later, after the publication of the paper) when the Google team checked the computation with their own simulators, they found that in a single case with 39 qubits the amplitudes were different, and 
also the estimated fidelity was lower than expected. 
Initially, the thought was that the (small) differences between the outcomes were caused by numerical differences between the simulators.
However, this was not the reason. 
It turned out that the problem was caused by the different calibration methods. (See Section \ref {s:cal}.) 
In the simulations coming from J\"ulich the researchers used circuits and measurements corresponding to the correct experiment,
but the Google team improved the parameters of the circuits a few days after running the experiment, and the J\"ulich team
used (for one circuit) an older version. This explains the discrepancy in the amplitudes and why the fidelity estimated by the J\"ulich team was lower for one of the circuits. 
In other words, the amplitudes were computed for precisely the same experiment, but the calibration, namely, the formalization of the
experiment as a quantum circuit (the description of the ``native" gates), was slightly different.

\subsection {Data from other sources}

It would be valuable to perform some of our statistical
analysis on data from other quantum computers and we are trying to obtain similar
data (descriptions of calibrated circuits, bitstrings, and amplitude files)
from other quantum computers, especially from the IBM quantum computer. The experiment closest to the Google one (that we know about) that was carried out on an IBM quantum computer is by Kim et al. \cite {Kim+21}. 
In addition, 
some of IBM's benchmarks are based on random circuits of some sort.
IBM, Google, NASA, and various other groups have powerful simulators that allow for the introduction of noise. Our statistical analysis can also be applied to probability distributions produced by these simulators.

\subsection {The USTC replications}

In June 2021 scientists from USTC \cite {Wu+21} described a close replication of the Google 2019 quantum supremacy experiment 
with 
56-qubit depth-20 quantum circuits on their {\it Zuchongzhi} quantum computer. Later, in September 2021, they described an improved experiment with 60-qubit circuits of depth 24 \cite {Zhu+22}. 
The team from USTC shared the description of the circuits (in Matlab) 
for each $n= 15,18,21,24,...,54, 56$, and it seems that they sampled (or shared) fewer bitstrings than Google did; for example, for $n=15$ they sampled 200K bitstrings, whereas 
the Google team sampled 500K. 
Carefully studying the data from these experiments (and perhaps asking for additional data) 
would be an interesting direction for further research. 

\subsection {Discussions in scientific blogs}

There were useful discussions in a few scientific blogs regarding the Google experiment,  
especially in Aaronson's ``Shtetl Optimized'' (SO) and Kalai's ``Combinatorics and More'' blogs. Reference \cite {blogs} contains links to useful blog discussions mainly at the time of the Google quantum supremacy announcement. For example, in a discussion in SO (December 2020) a commentator named ``Till'' asserted \cite {SO:Till} that the calibration process actually adjusted the definition of the circuit to the device, in other words, that the random circuits were generated with ``standard gates'' but the definition of 2-gates 
was modified in the calibration process. 
(See Section \ref {s:cal}.)
This was news to us and to several others who 
thought that the calibration process was a physical process
performed on the Google quantum computer. Aaronson's own conclusion from the discussion was: 
``So, my summary would be that yes, there’s a calibration phase,
and the 2-qubit gates used depend on the outcome of that
phase, but there’s still a
clear separation enforced between the calibration phase and the actual running of the QC.''
\section {Two proposals for future experiments} 

\subsection {A proposal for blind experiments}

In \cite {RSK22} we proposed the following protocol for an evaluation of Google’s experiment
(here we describe a small variant). First, Google will share the parameters of their calibration. Next, independent
scientists will prepare several programs (circuits) for Google's quantum computer to be run with about $n$ qubits, for which computing the sampling probabilities should be a task that takes several months (on a classical supercomputer). These programs will be sent to Google for implementation. Google will send back in a short time the implemented programs and large samples that they produce. This procedure is assumed to
preclude computation of the relevant amplitudes for the calibrated circuits. Using classical supercomputers the
scientists will take their time and compute the set of amplitudes for each calibrated circuit. They will then evaluate the relation between those amplitudes and the samples they received. Such a protocol is likely to be
relevant to other quantum supremacy demonstrations that are being pursued very actively these days. Overall, the Google team welcomed the idea of conducting blind experiments
in the future and agreed to our specific proposed protocol.

{\bf Remark:} During an Internet discussion (on Aaronson's blog in February 2020) 
Craig Gidney \cite {SO:Gidney}
asserted that various experiments 
that were initially considered computationally hard turned out to be easier than expected
and they served and could further serve as sort of blind tests for the experiment. 
The Google team made a similar comment in our email correspondence in January 2021:
``We did publish a lot of data that no one has been able to analyze yet.
I hope that eventually some of this data will be analyzed, which will be an interesting confirmation.
Analyzing 39 and 40 qubits has actually become easy in the meantime.'' 
These remarks motivated our request (h) in Section \ref {s:req}. As we mentioned, the vast progress in simulation techniques
 made it possible to examine the fidelity estimators and they have been verified 
by Kalachev, Panteleev, and Yung \cite {KPY21}.



\subsection {Testing calibration strategies by other groups on Google's data}

As a follow-up to our request for the calibration programs (item (i) in Section \ref{s:req}) 
that could not be met, 
we raised the possibility of sharing raw data gathered from 1-qubit and 2-qubit circuits that formed 
the input for the calibration programs needed to describe the ``native gates.'' This would give an
opportunity to other groups to test their own calibration
methods (without revealing Google's full proprietary calibration system). The Google team asserted that this might be possible in principle, but it would require considerable effort, and it is not clear whether the Google experiment data was kept. It might be
easier for Google to implement this proposal in more recent or even future Sycamore quantum computer experiments.
Given the central place of the calibration stage in NISQ experiments,
this direction could be valuable in its own right. 



\section {On the evaluation of the Google experiment}
In this section we discuss the overall evaluation of the Google experiment. 
We start with five central questions regarding the experiment and continue with a list of confirmations, refutations, concerns, and weaknesses of the experiment. Some issues discussed in this section are not directly related to the data and information gathering that we discussed in the previous sections, and we also refer to important works by other groups.

\subsection {Five central questions}

\label{s:5cc}

As we have already mentioned, several ingredients of the Google 
experiment represent major progress in our ability
to control quantum systems. Here are some
central problems that arise when we evaluate the Google experiment.  

\begin {itemize}

\item[a)] Were the sampling tasks achieved as claimed?
\end {itemize}
In our opinion, the findings of our paper \cite {RSK22} show that the answer is negative. A ``sampling task" in the traditional sense, namely,
reaching (approximately) a sample from some known distribution, was not achieved in the Google quantum supremacy experiment. The empirical distribution is quite different from  Google's basic noise model (\ref{e:gnm}) and our more detailed noise model yields only a small improvement. The Google team disagreed with our opinion and noted that they explicitly stated in the supplement \cite {Aru+19S} (around Equation (24)) that they  do not necessarily assume the basic noise model (\ref{e:gnm}).  

\begin {itemize}
\item
[b)] Are the statistical tools for estimating the fidelity satisfactory?
\end {itemize}

In our study \cite {RSK22} we found Google’s statistical framework to be very 
sound. In particular,  the ${\mathcal F}_{XEB}$ estimator is good under the Google noise model and is quite robust to changes in the model. (We offered some technical improvements to Google's estimator mainly when the number of qubits is small.)
Gao et al.\cite {Gao+21} noted that there could be some systematic gap between the ${\mathcal F}_{XEB}$ estimator and the fidelity for noisy quantum circuits but estimated the gap to be small. 
We note that it might be legitimate to base quantum supremacy claims on the computational hardness of sampling with a high value of ${\mathcal F}_{XEB}$, without referring to the question of whether a specific sampling task was achieved or whether
${\mathcal F}_{XEB}$ genuinely estimated the fidelity. (See \cite {AarGun19,Gao+21}.)

\begin {itemize}
\item
 [c)] Are the claims regarding the estimated fidelity of the samples valid?

  \end {itemize}

  The Google team claimed to achieve samples for random circuit sampling (full circuits) with $m=14$ and $n=12$ with estimated fidelity 0.3694; for $m=14$ and $n=22$ with estimated fidelity 0.165; and for  $m=14$ and 
  $n=32$ with estimated fidelity 0.071.  
  Here, the question is to what extent this achievement can be regarded as solid. 
  We note that the specific computations for the ${\mathcal F}_{XEB}$ value of the samples, given the 
  description of the (calibrated) circuits, were verified. Thus, the remaining question is whether the overall claim is sound.

\begin {itemize}
\item[d)] 

Are the claims regarding the predictive power of the a priori fidelity estimations valid? (Here we refer to 
Formula (\ref {e:77}).)

\end {itemize}

  The a priori predictions based on Formula (\ref {e:77}) (Formula (77) in \cite {Aru+19S}) for $m=14$ and $n=12,22,32$
  are $0.386, 0.1554$, and $0.062$, respectively. 
  Question d) is whether we have solid evidence for the claim that 
  the simple Formula (77) (and the statistical independence assumptions on which it rests) provides an accurate prediction of the ${\mathcal F}_{XEB}$ estimator. Here, the specific computations were not checked (the individual fidelities were not shared, see item (f) in Section \ref {s:ddd}), but they are supported by our approximate computations based on averaged values of the fidelities. 
  As in the previous item, the remaining question is about the soundness of Google's overall claim and here
  the excellent predictive power of the a priori fidelity estimation raised concerns about it. (See item 10 in Section \ref {s:crcw} below.)

\begin {itemize}
  \item
[e)] Are the claims regarding the classical difficulty of the sampling task correct?

\end {itemize}

Regarding Google's quantum supremacy claim, a very short summary is that by now classical algorithms
are ten orders of magnitude faster than those used in the Google paper, and largely refute  Google’s fantastic claims. 
(See, e.g., \cite{PanZha21,PCZ22,ZSW20,KPY21,KPZY21,PGN+19}.)
The Google paper claims that their ultimate task that required 200 seconds on the quantum
computer would require 10,000 years on a classical supercomputer. 
With the new algorithms the task can be completed in a matter of seconds.

A few days after the publication of the Google quantum supremacy paper researchers from IBM \cite{PGN+19} achieved a theoretical improvement of six orders of magnitude (albeit on a more powerful supercomputer). In response, 
the
Google team pointed out that their paper \cite {Aru+19} did anticipate some progress in classical algorithms: ``We expect that lower simulation costs than reported here will eventually be achieved, but we also expect that they will be consistently outpaced by hardware improvements on larger quantum processors." As a matter of fact, the Google team welcomed better algorithms and wrote in \cite {Aru+19} that the bitstring samples from all circuits had been archived \cite {Aru+19D} 
``to encourage development and testing of more advanced verification algorithms."

While weakening Google's quantum supremacy claims, the progress in classical algorithms has also led to major support for Google's claims on items c) and d).
In a recent work \cite {KPY21},  
Kalachev, Panteleev, and Yung were able to compute the fidelity
of samples that Google's algorithms (from 2019) were too slow to handle. The fidelity values 
that the authors computed
agree perfectly with Google's Formula (77) predictions. 
This lends very strong support to an affirmative answer to items c) and d).
Another major support for items c) and d) is two replications by a team from USTC of the Google experiment \cite {Wu+21,Zhu+22}. 
Yet more support for items c) and d) came from our own study. In \cite {RSK22} we offered (as a ``sanity test'')
an alternative way to estimate the fidelity
based on a ``secondary'' component of the theoretical distributions that arises from readout errors. In a subsequent paper \cite {RSK22+},
we checked these alternative estimators for the fidelity and found a good match with the ${\mathcal F}_{XEB}$ estimators for $n \le 30$. (See Figure 2.) This
work nicely implements Fourier tools.

\begin{figure}
\centering
\includegraphics[scale=0.8]{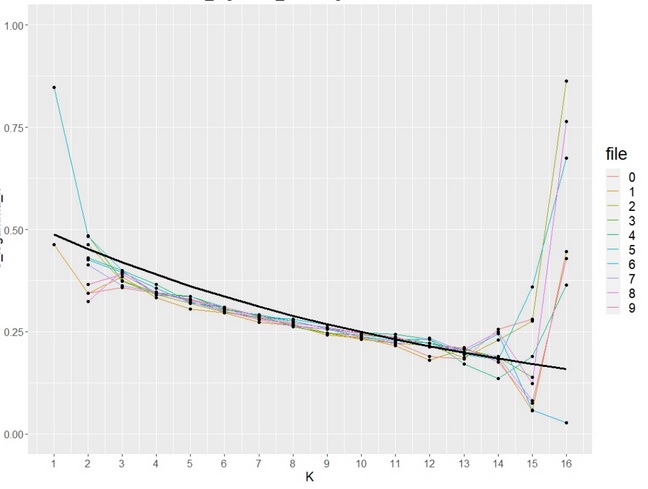}
\caption{{\it The decay of Fourier–Walsh contributions as predicted by the theory (bold black curve)(\cite{BSN17,RSK22+}) and as demonstrated by samples of a 16-qubit Sycamore computer (ten random circuits).}}
\label{fig:2}
\end{figure}

\subsection {Confirmations, refutations, concerns, and weaknesses}
\label {s:crcw}

Here is a list, without much commentary, of the weaknesses, concerns, confirmations, and refutations of the
Google quantum supremacy experiment.

\subsection *{A) Confirmations}
\begin {enumerate}
\item
The computations of the fidelity estimator and related computations
reported in the Google paper
were confirmed by us up to our own computational limits (and by others). 
\item
A major confirmation of the Google fidelity predictions in the ``supremacy region'' was achieved by Kalachev, Panteleev, and Yung  \cite {KPY21}. This lends
strong support to the reliability of the Google experiment. Another 
major confirmation of the ultimate $n=53, m=20$ quantum supremacy experiment was achieved by 
Liu et al. \cite {Liu+22}.

\item 
Our readout error study and the related study of the Fourier--Walsh expansion of the empirical data largely supports (for $n \le 30$) Google's experimental
claims and the experiment's reliability (work in progress \cite {RSK22+}). (See Figure \ref{fig:2}.) 

\item
 Successful replications were announced and published by a group from USTC \cite {Wu+21,Zhu+22}.
\item
The Google fidelity estimator and statistical methodology are overall very sound \cite {RSK22}.
\end {enumerate}

\subsection *{B) Refutations} 
\begin {enumerate}
\item [6.] Google's quantum supremacy claims were largely (but not fully) refuted \cite{PanZha21,PCZ22,ZSW20,KPY21,KPZY21,PGN+19}. (See Section \ref {s:5cc} above.)

\item
[7.] The data does not fit the Google noise model \cite {RSK22} or any other specific model. (See Section \ref {s:5cc} above.); However, the Google team disagrees with this point. 
\end {enumerate}

\subsection *{AB) Mixed confirmation/refutation}
\begin {enumerate}
\item
[8.] The boson sampling quantum supremacy experiments provide inconclusive evidence. They were regarded as independent confirmations of quantum supremacy using a very different quantum device; however, the computational hardness claims are in tension with old \cite {KalKin14} and new \cite {PopRub21,Vil+21} research. 
\end {enumerate}

\subsection *{C) Concerns}
\begin {enumerate}

\item [9.] The Google experiments represent amazing leaps in our ability to control quantum devices. These breakthroughs are surprising and there isn't a clear technological reason for them.\footnote{The Google quantum supremacy experiment has often been compared to IBM's ``Deep Blue" victory over world chess champion Kasparov; in that case, however, progress was gradual and earlier chess programs could already beat chess grandmasters.}
\end {enumerate}
We note that here we are referring to the ability to perform random circuit sampling experiments. The remarkable progress in building the hardware itself (by several groups in academia and industry including the Google team) was gradual.

\begin {enumerate}
\item [10.] The predictive power of the a priori fidelity estimation based on the fidelities of the individual components of the quantum computer (Formula (77) in \cite {Aru+19S})  raised concerns about reliability of the Google experiment. This matter is raised and discussed in \cite {Kal22} and is also mentioned in \cite {RSK22}.
\end {enumerate}

The concern regarding the excellent prediction of the ${\cal F}_{XEB}$ fidelity estimator based on the fidelities of individual components, appears to be the strongest challenge to date to the reliability of the Google experiment and the USTC replications. There are two important remarks regarding this concern. First, the Google team and other
researchers regard this
predictive power as one of the main achievements of the Google experiment.
Second, \cite {KPY21} confirmed these predictions
for many circuits for which the Google team could not even compute the amplitudes.
(This is a sort of ``blind test'' verification.)

\begin {enumerate}
\item [11.] Since the Google's noise model does not describe the empirical data, the fidelity ${\cal F}_{XEB}$ estimator may not correctly capture the fidelity of the quantum system. (See, e.g.,  \cite{Gao+21}.)

\item [12.] Some aspects of the calibration process may seem overly surprising. 

\item [13.] There are no sufficient replications of the supremacy experiment even in the 12--40 qubit range (either by the Google team or by other teams).  

\end {enumerate}

\subsection *{D) Weaknesses} 
\begin {enumerate}

\item [14.] While the Google experiment was characterized as performing a  sampling task, the empirical distribution was not compared to the noise model, and the experimental samples were too small to allow such a comparison (directly) when $n>14$.

\item [15.] Google's quantum supremacy claim is based on a comparison with a single classical algorithm. 
The Google team itself developed better algorithms for related computational tasks.

When the Google team found an improved type of algorithm (around May 2019) they
switched to a different class of circuits to which the new algorithm could not not be applied without
sufficient evidence that similar improvements could not be made for them as well. (Later, as we mentioned, improvements were also made to the new class of circuits.)

\item [16.] The results were not presented to the academic community before publication (e.g., as an arXiv publication), and 
careful review of the experiment prior to publication may have been  
insufficient. 

\item [17.] In view of the rigid nature of the experimental circuits, the reference to a ``programmable quantum computer'' in the paper's title is unjustified.

\item [18.] The notion of quantum supremacy as a scientific claim is problematic. This concern was raised by Borcherds \cite {Bor21} 
and several other scientists. 




\item [19.] The main fidelity estimator ${\cal F}_{XEB}$ is not unbiased and can be improved \cite {RSK22}.


\item [20.] Changing to a new pattern {\bf ABCDCDAB} weakened the power of the extrapolation argument. For the new pattern no experiments for
circuits with fewer than 53 qubits were conducted.
(This was a weakness in the planning of the experiment although some of the fidelity predictions are now confirmed \cite {KPY21}.)

\item [21.] Improvements to the calibration process were carried out at the same time as 
the experiment was being run. The description of the calibration process in Google's popular video \cite{Google:video19} may be in tension with the claim that there was a clear separation between the calibration stage and the experiment itself. 

\item [22.] The calibration process itself further weakens the claim for a ``programmable quantum computer.'' (This concern was raised
by a commentator ``Till" in an Internet discussion \cite {SO:Till}; it was also raised and discussed even earlier by Gidney  from the Google team \cite {SO:Gidney2}).) 

\item [23.] The effect of the calibration process is very large (it is not a ``fine tuning'' of some sort).

\item [24.] The precise programs for the calibration process are commercial secrets. 
(However, the main innovation in the calibration was the use of XEB,
and that code is open-sourced.)

\item [25.] The levels of transparency and of documentation of the Google experiment seem insufficient.   
\end {enumerate}

Of the items presented above some are controversial, some are rather minor, and many are interrelated. Items (15) and (16) are a priori weaknesses that turned out to be crucial (as seen by refutation (6) above). The recent confirmations (2) indicate that weakness (20) may have no bearing on the predictions for large circuits, but we still regard it as a weakness in the planning of the experiment. (The Google team disagrees with us.) 

A recent remarkable work of Gao, Kalinowski, Chou, Lukin, Barak, and Choi \cite {Gao+21} is relevant to some of the concerns discussed above. Analysis based on \cite {Gao+21} strengthens concern (10) and casts some doubt on our confirmation (3). For a random quantum circuit $C$, Gao et al.'s paper describes a method for producing samples with high linear cross-entropy value, without computing the probabilities described by $C$.

We also note that the expectations expressed in the Google paper \cite {Aru+19}  
seem overly optimistic: 
\begin {quotation}
``Quantum processors have thus reached the regime of quantum supremacy. We expect that their computational power will continue to grow at a double-exponential rate: the classical cost of simulating a quantum circuit increases exponentially with computational volume, and hardware improvements will probably follow a quantum-processor equivalent of Moore’s law, doubling this computational volume every few years." 
\end {quotation}

There is no evidence for doubly exponential growth (or even slower growth) of computational power, and there are theoretical reasons (\cite {Kal20:p,ZBL20,Gao+21}) to doubt that these expectations will be fulfilled.












\section {Where we are}

Now that the data-gathering stage of the Google experiment needed 
for our study has largely come to an end,  
we can present a brief summary of where we are. 

1. We compared the empirical distribution with Google's noise model and with other noise models. (Some of this analysis is already discussed in \cite {RSK22}, and a few other findings were discussed with the Google team.)

2. 
In \cite {RSK22} we offered  an alternative way (that could serve as a ``sanity test") to estimate the fidelity based on a linear cross-entropy test for a ``secondary'' component of the theoretical distributions that arises from readout errors. In a subsequent work in progress \cite {RSK22+} we used Fourier methods to check this alternative estimator for the fidelity, and found a good match with the primary linear cross-entropy fidelity estimators for $n \le 30$. 


3. We tried to understand various aspects of the calibration process and carried out some statistical analysis for this purpose.


4. We studied various aspects of the predictive power of Google's Formula (77). The remarkable predictive power of Formula (77) in \cite {Aru+19S} is regarded as an important scientific discovery on its own, and we regard it as a central concern.  

5. We are extending some of our statistical study to the new data from June 2022,
and especially to the data for the patch circuits. There is also some remaining analysis of the earlier data. We plan to extend our statistical study to data coming from other NISQ computers and data from simulators that incorporate the effect of noise.

\bigskip
\subsection *{Acknowledgements}
Research supported by ERC grant 834735. We thank many colleagues including members of the Google 
team for helpful discussions, Carsten Voelkmann for many corrections and helpful suggestions, and Mike Borns for thorough editing.


\noindent
{\small Gil Kalai,  Hebrew University of Jerusalem and Reichman University, {\tt gil.kalai@gmail.com}

\noindent
Yosef Rinott,  Hebrew University of Jerusalem, {\tt yosef.rinott@mail.huji.ac.il}

\noindent
Tomer Shoham,  Hebrew University of Jerusalem, {\tt tomer.shohamm@gmail.com}
}


\begin{thebibliography}{99}


\bibitem {Aar19} S. Aaronson, Why Google’s quantum supremacy milestone matters, Opinion, {\it New York Times}, Oct. 30, 2019.



\bibitem {AarGun19}S. Aaronson and S. Gunn, On the classical hardness of spoofing linear cross-entropy benchmarking, 2019, arXiv:1910.12085.



\bibitem {Aru+19} 
F. Arute et al., Quantum supremacy using a programmable superconducting processor, 
{\it Nature,} 574 (2019), 505--510.

\bibitem {Aru+19S}
  F. Arute et al.,
Supplementary information for ``Quantum supremacy using a programmable 
superconducting processor,'' 2019, arXiv:1910.11333. 

\bibitem {Aru+19D}
  F. Arute et al.,
Supplementary data for ``Quantum supremacy using a programmable 
superconducting processor," 2019, \href {https://datadryad.org/stash/dataset/doi:10.5061/dryad.k6t1rj8} {https://datadryad.org/stash/dataset/doi:10.5061/dryad.k6t1rj8}. 



\bibitem {BISN17} S. Boixo, S. V. Isakov, V. N. Smelyanskiy, and H. Neven, Simulation of low-depth quantum circuits as complex undirected graphical models, 2017, arXiv:1712.05384.


\bibitem {BSN17} S. Boixo, V. N. Smelyanskiy, and H. Neven,
Fourier analysis of sampling from noisy chaotic quantum circuits, 2017, arXiv:1708.01875.

\bibitem{Bor21} R. E. Borcherds, \href {https://youtu.be/sFhhQRxWTIM}{The teapot test for quantum computers }, a videotaped lecture, 2021, URL: https://youtu.be/sFhhQRxWTIM. 


\bibitem{Gao+21}
X. Gao, M. Kalinowski, C.-N. Chou, M. D. Lukin, B. Barak, and S. Choi, 
Limitations of linear cross-entropy as a measure for quantum advantage, 2021, arXiv:2112.01657.

\bibitem {Google:video19} Google, \href {https://youtu.be/-ZNEzzDcllU}{Demonstrating Quantum Supremacy },  2019,
URL: https://youtu.be/-ZNEzzDcllU.

\bibitem {Ira19} S. Irani (Moderator), \href {https://youtu.be/\_Yb7uIGBynU}{Supremacy Panel}, Participants: D. Aharonov, B. Barak, A. Bouland, G. Kalai,
S. Aaronson, S. Boixo, and U. Vazirani, 2019, 
 URL: https://youtu.be/\_Yb7uIGBynU.
\bibitem {KPY21} G. Kalachev, P. Panteleev, and M.-H. Yung, 
Multi-tensor contraction for XEB verification of quantum
circuits, 2021, 
arXiv:2108.05665.

\bibitem {KPZY21} 
G. Kalachev, P. Panteleev, P. F. Zhou, and M.-H. Yung, 
Classical sampling of random quantum circuits with bounded fidelity, 2021, arXiv:2112.15083.




\bibitem {Kal20:p} G. Kalai, The argument against quantum computers, in: M. Hemmo, and O. Shenker (eds.) 
{\it Quantum, Probability, Logic: Itamar Pitowsky's Work and Influence}, 
Springer (2020) ,pp. 399--422, arXiv:1908.02499.

\bibitem {Kal20} G. Kalai, \href {https://youtu.be/p18P1y8GD9U} {The Google quantum supremacy demo }, 2019, URL: https://youtu.be/p18P1y8GD9U.

\bibitem {Kal22} G. Kalai, The argument against quantum computers, the quantum laws of nature, and
Google’s supremacy claims, in: {\it The Intercontinental Academia Laws: Rigidity and Dynamics} (M.
J. Hannon and E. Z. Rabinovici (eds.)) Proceedings of the ICA Workshops 2018 \& 2019, Singapore and
Birmingham, World Scientific (to appear, 2023). arXiv:2008.05188.

\bibitem{KalKin14} G. Kalai and G. Kindler, Gaussian noise sensitivity
and BosonSampling, 2014, arXiv:1409.3093.


\bibitem {Kim+21} J.-S. Kim, L. S. Bishop, A. D. Corcoles, S. Merkel, J. A. Smolin, S. Sheldon, Hardware-efficient random circuits to classify noise in a multi-qubit system, 
{\it Physical Review A}, 104 (2021), 022609. 







\bibitem {Liu+22} Y. Liu et al., Validating quantum-supremacy experiments with exact and fast tensor network contraction, 2022, arXiv:2212.04749. 


\bibitem {Nei+18} 
C. Neill et al.,
A blueprint for demonstrating quantum supremacy with superconducting qubits, {\it Science}, 360 (2018), 195--199.



\bibitem {PCZ22} 
F. Pan, K. Chen, and P. Zhang, 
Solving the sampling problem of the Sycamore quantum supremacy circuits, 
{\it Physical Review Letters}, 129 (2022), 090502.

\bibitem {PanZha21} 
F. Pan and P. Zhang, Simulating the
Sycamore quantum supremacy circuits, 2021, 
arXiv:2103.03074


\bibitem {PGN+19}
E. Pednault, J. A. Gunnels, G. Nannicini, L. Horesh, and R. Wisnieff, 
Leveraging secondary storage to simulate deep 54-qubit Sycamore circuits, 2019, 
arXiv:1910.09534.



\bibitem {PopRub21} A. S. Popova and A. N. Rubtsov, Cracking the quantum advantage threshold for Gaussian boson sampling,
  2021, arXiv:2106.01445.





\bibitem {RSK22} Y. Rinott, T. Shoham, and G. Kalai, Statistical aspects of the quantum supremacy
demonstration, {\it Statistical Science}, 37 (2022), 322--347.

\bibitem {RSK22+} Y. Rinott, T. Shoham, and G. Kalai, Quantum advantage demonstrations via random circuit sampling: Fourier expansion and statistics, {\it work in progress}.


\bibitem{SO:Boixo} {\it Shtetl Optimized}, \href {https://scottaaronson.blog/?p=5159#comment-1870651}
{A comment by S. Boixo } , Dec. 2020,
URL: https://scottaaronson.blog/?p=5159\#comment-1870651.

\bibitem{SO:Gidney2} {\it Shtetl Optimized}, \href {https://scottaaronson.blog/?p=4372#comment-1822373}{A comment by C. Gidney }, Nov. 2019,
URL: https://scottaaronson.blog/?p=4372\#comment-1822373.

\bibitem{SO:Gidney} {\it Shtetl Optimized}, \href {https://scottaaronson.blog/?p=4608#comment-1830545}{A comment by C. Gidney }, Feb. 2020,
URL:  https://scottaaronson.blog/?p=4608\#comment-1830545.



\bibitem {SO:Till} {\it Shtetl Optimized}, \href {https://scottaaronson.blog/?p=5159#comment-1869118}{A comment by ``Till" }, Dec. 2020, URL:  https://scottaaronson.blog/?p=5159\#comment-1869118.





\bibitem {Vil+21}
  B. Villalonga, M. Yuezhen Niu, L. Li, H. Neven, J. C. Platt, V. N. Smelyanskiy, and S. Boixo,
Efficient approximation of experimental Gaussian boson sampling, 2021, arXiv:2109.11525.








\bibitem {Wu+21} Y. Wu et al., Strong quantum computational advantage using a superconducting quantum processor. {\it Physical Review Letters}, 127 (2021), 180501. 

\bibitem {Zho+20} H.-S. Zhong et al., Quantum computational advantage using photons, 
{\it Science}, 370 (2020),
1460–1463.

\bibitem {ZSW20} 
Y. Zhou, E. M. Stoudenmire, and X. Waintal, 
What limits the simulation of quantum computers?, 2020,  
arXiv:2002.07730.

\bibitem {Zhu+22} Q. Zhu et al., Quantum computational advantage via 60-qubit 24-cycle random circuit sampling, {\it Science Bulletin}, 67 (2022), 240--245.
arXiv:2109.03494.


\bibitem {ZBL20}  A. Zlokapa, S. Boixo, and D. Lidar, Boundaries of quantum supremacy via random circuit sampling, 2020, 
arXiv:2005.02464.

\bibitem{blogs}Various links to useful discussions over scientific blogs of the Google's 2019 paper and related developments. ({\bf SO} stands for Shtetl Optimized, Scott Aaronson's blog; {\bf CaM} stands for Combinatorics and More, Gil Kalai's blog; {\bf WoT} stands for Windows on Theory, Boaz Barak's blog.

\begin {enumerate}
\item
\href {https://scottaaronson.blog/?p=4317} {Scott’s supreme quantum supremacy FAQ! } {\bf SO}, Sep. 23, 2019.
https://scottaaronson.blog/?p=4317.
\item 
\href {https://gilkalai.wordpress.com/2019/09/23/quantum-computers-amazing-progress-google-ibm-and-extraordinary-but-probably-false-supremacy-claims-google/}
{Quantum computers: Amazing progress (Google \& IBM), and extraordinary but probably false supremacy claims (Google).} {\bf CaM}, Sep. 23, 2019.
https://gilkalai.wordpress.com/2019/09/23/quantum-computers-amazing-progress-google-ibm-and-extraordinary-but-probably-false-supremacy-claims-google/
\item
\href
{https://gilkalai.wordpress.com/2019/10/13/the-story-of-poincare-and-his-friend-the-baker/}
{The story of Poincaré and his friend the baker }, {\bf CaM}, Oct. 13, 2019.
https://gilkalai.wordpress.com/2019/10/13/the-story-of-poincare-and-his-friend-the-baker/
\item
\href {https://scottaaronson.blog/?p=4372}  {Quantum supremacy: The gloves are off }, {\bf SO}, Oct. 23, 2019/  https://scottaaronson.blog/?p=4372.
\item
\href{https://windowsontheory.org/2019/10/24/boazs-inferior-classical-inferiority-faq/}
{Boaz’s inferior classical inferiority FAQ }, {\bf WoT}, Oct. 24, 2021.
https://windowsontheory.org/2019/10/24/boazs-inferior-classical-inferiority-faq/.
\item       
\href{https://gilkalai.wordpress.com/2019/11/13/gils-collegial-quantum-supremacy-skepticism-faq/}
{Gil’s collegial quantum supremacy skepticism FAQ }, {\bf CaM} Nov. 13, 2019,
https://gilkalai.wordpress.com/2019/11/13/gils-collegial-quantum-supremacy-skepticism-faq/.
\item
 \href {https://scottaaronson.blog/?p=4608}{My “Quantum supremacy: Skeptics were wrong”  2020 world speaking tour }, 
 {\bf SO}, Feb. 17, 2020 https://scottaaronson.blog/?p=4608.
\item
\href {https://scottaaronson.blog/?p=5122}
{Quantum supremacy, now with BosonSampling },
{\bf SO}, Dec. 3, 2020,
https://scottaaronson.blog/?p=5122.
\item
\href
{https://scottaaronson.blog/?p=5159}
{Chinese BosonSampling experiment: The gloves are off },
{\bf SO}, Dec. 16, 2020. 
https://scottaaronson.blog/?p=5159.

\end {enumerate}




\end{thebibliography}
\end {document}